\documentclass[a4paper,cits,proceedings]{PoS}
\usepackage{amssymb,amsmath}
\newcommand{\be}{\begin{equation}}
\newcommand{\ee}{\end{equation}}
\newcommand{\bea}{\begin{eqnarray}}
\newcommand{\eea}{\end{eqnarray}}

\newcommand{\MS}{\overline{\rm MS}}

\title{The static quark self-energy at large orders from NSPT}

\ShortTitle{The static quark self-energy at large orders from NSPT}

\author{Gunnar S.\ Bali, \speaker{Clemens~Bauer}\\
Institut f\"ur Theoretische Physik, Universit\"at Regensburg\\
D-93040 Regensburg, Germany\\
        E-mail: \email{gunnar.bali@ur.de}\\
        \hspace*{1.2cm}\email{clemens.bauer@physik.uni-regensburg.de}}

\author{Antonio Pineda\\
Grup de F\'{\i}sica Te\`orica, Universitat
Aut\`onoma de Barcelona,\\E-08193 Bellaterra, Barcelona, Spain\\
E-mail: \email{AntonioMiguel.Pineda@uab.es}}
\abstract{
Using Numerical Stochastic Perturbation Theory (NSPT), we calculate the static self-energy of $\mathrm{SU(3)}$ gauge
theory up to order $\alpha^{20}$. Simulations on a large set of different lattice volumes allow for a careful treatment of finite size effects. 
The resulting infinite volume perturbative series of the static self-energy is in remarkable agreement with the predicted asymptotic behaviour of high order expansions, 
namely with a factorial growth of perturbative coefficients known as renormalon.
}

\FullConference{XXIX International Symposium on Lattice Field Theory \\
		 July 10 -- 16 2011\\
		 Squaw Valley, Lake Tahoe, California}

\begin{document}

\section{Motivation}
Since the early days of quantum field theories, it has been argued that their perturbative weak coupling expansions are asymptotic~\cite{Dyson:1952tj}.
Usually, the asymptotic behaviour is identified using semiclassical methods such as instantons. Asymptotically free theories with marginal operators constitute
a special case. 
Examples of such theories are four-dimensional non-Abelian gauge theories or the two-dimensional $\mathrm{O}(N)$ model.
For these the structure of the operator product expansion (OPE) is believed to give rise to a specific pattern of asymptotic divergence 
 known as (infrared) renormalon~\cite{Hooft,Beneke:1998ui} that cannot be obtained using standard semiclassical methods. 
In the two-dimensional
$\mathrm{O}(N)$ model renormalon effects were found (though suppressed) in an 
explicit calculation in the large $N$ expansion~\cite{David:1982qv}. For four-dimensional 
non-Abelian gauge theories no such proof exists. 
On the contrary, the possibility that the renormalon either does not 
exist or is very small has been raised in recent years, see
e.g.~\cite{Suslov:2005zi,Zakharov:2010tx}.

At present, the only way to unambiguously settle this issue is by performing weak coupling expansions to sufficiently high orders in the strong coupling
parameter $\alpha$. In particular,
in $\mathrm{SU(3)}$ gauge theory, the problem has been addressed by
simulations in lattice regularization, using
Numerical Stochastic Perturbation Theory (NSPT)~\cite{DRMMOLatt94,DRMMO94,DR0}.
These simulations, that mainly focused on computing the plaquette
and small Wilson loops to perturbative orders as high as
$\alpha^{20}$~\cite{Horsley:2010af}, so far had no success in  
uncovering a renormalon.

The situation radically changed recently  when, for the first time, accurate agreement between the explicit computation of perturbative
coefficients and the expectation from the renormalon analysis was reported \cite{Bauer:2011ws}.
Here we discuss some aspects of this analysis where
the comparison was performed using another observable, the static
self-energy. This, on theoretical grounds, is far more suited for a
renormalon study. The simulations were conducted on a large set of
different lattice volumes, for orders as high as $\alpha^{20}$, and
finite size effects were carefully investigated. 

\section{Renormalon Theory}
In QCD, the expansion of a generic observable~$K$ as a power series in the coupling~$\alpha$,
\begin{equation}
K=\sum_n k_n\alpha^n,
\end{equation}
is believed not to be convergent but at best to be asymptotic. 
Technically, renormalons appear as singularities in the Borel plane, 
making a Borel summation impossible. The way in which
the series~$K$ diverges is closely tied to the OPE, 
commonly giving rise to terms~$k_n\sim a_d^nn!$, where $a_d$ is a constant. For small orders~$n$, 
the successive contributions $k_n\alpha^n$ to the series
reduce in size down to a minimum at $n_0\approx 1/(|a_d|\alpha)$. 
The series should be truncated at $n_0$ and inevitably one has to deal with an ambiguity of the order
of the minimum term,
$k_{n_0}\alpha^{n_0}\sim\exp[-1/(|a_d|\alpha)]$.

The OPE allows one to isolate the short 
(given by Wilson coefficients $C_i(q,\mu)$, where $i$ is the dimension) and long distance effects 
(matrix elements $\langle O_i(\mu,\Lambda)\rangle$) of an observable $R(q,\Lambda)$ by means of a factorization
\begin{equation}
\label{eq:opef}
R=C_0(q,\mu)\langle O_0(\mu,\Lambda)\rangle
+C_d(q,\mu)\langle O_d(\mu,\Lambda)\rangle\!\!\left(\frac{\Lambda}{q}\right)^d\!\!+\cdots\,.
\end{equation}
Here, $\mu$ is the factorization scale separating the perturbative and low momentum scales~$q$ and $\Lambda$ from each other: $q\gg\mu\gg\Lambda$.

For the plaquette, $\langle O_0\rangle=1$, followed by the dimension
$d=4$ gluon condensate as the next higher dimensional non-vanishing operator. 
The lower bound for the accuracy of expanding $C_0$ is of $\mathcal{O}(\Lambda^4/q^4) \sim k_{n_0}\alpha^{n_0}$, 
the size of the minimum term, as can be seen from
\begin{equation}
\label{eq:renormal}
\left(\frac{\Lambda}{q}\right)^d\simeq\exp\left(-\frac{1}{|a_d|\alpha}\right)\,,\quad\mbox{where}\quad |a_d|=\frac{\beta_0}{2\pi d}\,,\quad \beta_0=11\,.
\end{equation}
The (infrared) renormalon ambiguity 
of this perturbative series cancels that of the next order non-perturbative matrix element in eq.~(\ref{eq:opef}). 
Hence, the physical quantity $R$ is well-defined. It follows that
series expansions with the smallest $d$ 
(and therefore $n_0$) are those
in which one should be able to
detect renormalons most easily (for $d=1$ four times ``faster''
than in the plaquette case).
Among such candidates are the pole mass or the associated self-energy of a static source,
\begin{equation}
\label{deltam}
\delta m=\frac{1}{a}\sum_{n\geq 0}c_n\alpha^{n+1}(1/a)\,.
\end{equation}
$a^{-1}$, the inverse lattice spacing, serves as a UV-regulator.
The large $n$ behaviour of the coefficients $c_n$ reads
\begin{align}
\label{generalm}
c_n \stackrel{n\rightarrow\infty}{=}
N_{m}\,\left(\frac{\beta_0}{2\pi}\right)^n
\,\frac{\Gamma(n+1+b)}{
\Gamma(1+b)}
\left(
1+\frac{b}{(n+b)}s_1+ \cdots
\right)\,,
\end{align}
where the coefficients $b$ and $s_1$ can be found in \cite{Beneke:1994rs} and $N_m$ is a 
normalization constant that cancels when constructing ratios
\be
\label{cnratio}
\frac{c_{n}}{c_{n-1}}\frac{1}{n} =\frac{\beta_0}{2\pi}
\left[1 +\frac{b}{n} - (1-b\,s_1)\frac{b\,s_1}{n^2}
+\mathcal{O}\left(\frac{1}{n^3}\right)
\right]
\,.
\ee
\section{Numerical Simulation}
The simulations are performed within the framework of 
NSPT~\cite{DRMMOLatt94,DRMMO94,DR0} that allows for a direct determination of perturbative series
coefficients. A
key ingredient of NSPT is the Langevin update, for which we 
employ the $\mathcal{O}(\Delta\tau^2)$ integrator described in
\cite{Torrero:08,BBP}.
Its convergence behaviour regarding the time step $\Delta\tau$
is remarkably flat, so that we can circumvent a $\Delta\tau$-extrapolation
and work at a fixed value $\Delta\tau = 0.05$. We implement periodic
boundary conditions in time and 
twisted boundary conditions~\cite{tHooft79,Parisi83,Luscher86,Arroyo88}
in the three spatial directions, completely
eliminating zero modes. 
Twists in two directions would have achieved
this, too~\cite{Bauer:2010jb}, but with
reduced numerical stability.

Employing the Wilson gauge action, we calculate the temporal Polyakov line
\begin{equation}
L^{(R)}(N_S,N_T)=\frac{1}{N_S^3}\sum_{\mathbf n}\frac{1}{d_R}
\mathrm{tr}\left[\prod_{n_4=0}^{N_T-1}U^R_4(n)\right]\,.
\end{equation}
on hypercubic volumes with $N_S$ and $N_T$ lattice points in spatial
and temporal directions, respectively. 
$R$ denotes the representation of the link $U^R_{\mu}(n)$, of dimension $d_R$,
and we study both, the triplet and the octet cases.
We repeat the measurements using stout-smeared~\cite{Morningstar:2003gk}
(smearing parameter $\rho=1/6$) links in the temporal direction. 
Altogether, this amounts to four distinct self-energies (allowing for checks of Casimir scaling and universality) 
whose perturbative coefficients~$c_n$ are linked to the Polyakov line via
 \begin{equation}
\label{eq:defP}
P(N_S,N_T)
=-\frac{\ln\langle L(N_S,N_T)\rangle}{aN_T}\
\stackrel{N_S,N_T\rightarrow\infty}{\longrightarrow} \delta m
\,.
\end{equation}
\begin{figure}[t!]
\centerline{\includegraphics[clip,width=.93\textwidth]{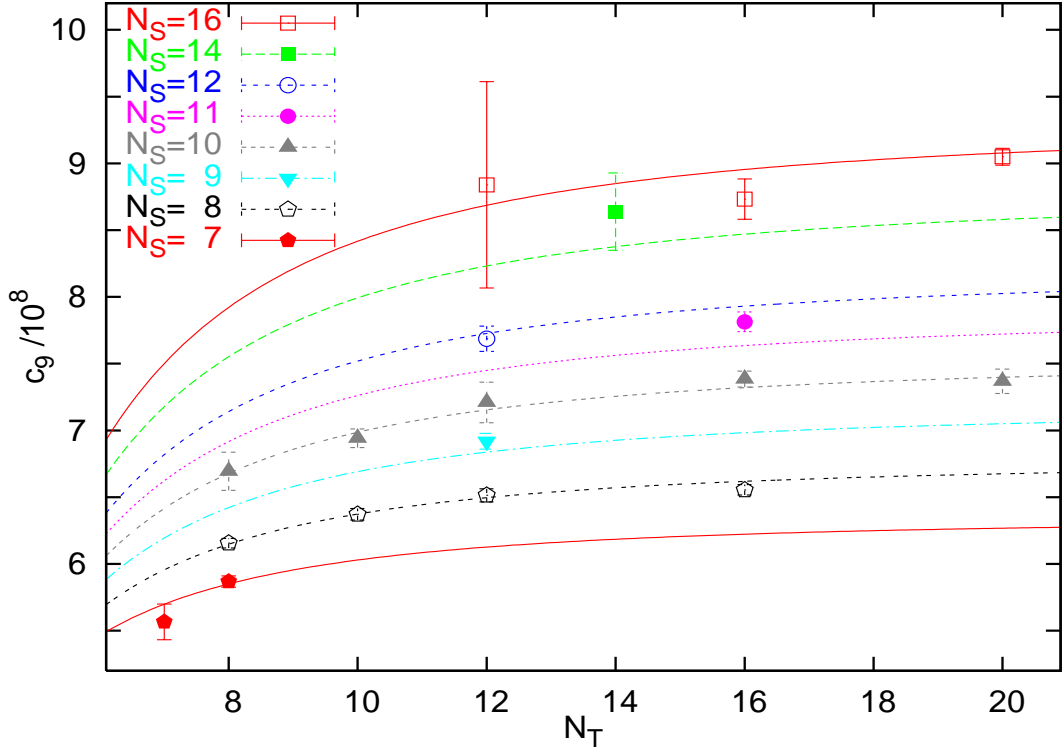}}
\caption{Comparison between the global fit and data for $n=9$
($\mathcal{O}(\alpha^{10})$).
\label{fig:fit}}
\end{figure}
The dependence on $N_T$ and $N_S$ can be parameterized \cite{Bauer:2011ws} as
\begin{align}\nonumber
aP&=\sum_{n\geq 0}
\left[
c_n\alpha^{n+1}\!\left(a^{-1}\right)
-\frac{f_n}{N_S}\alpha^{n+1}\!\!\left(\left(aN_S\right)^{-1}\right)+\cdots
\right]\\\nonumber
&=\sum_{n\geq 0}
\left[c_n+\Delta_n^{(1)}(N_S)+\Delta_n^{(2)}(N_S,N_T)\right]\!\alpha^{n+1}\!\!\left(a^{-1}\right)\,,
\\\label{eq:dlo}
\Delta_n^{(1)}&=-\frac{1}{N_S}\left[f_n+\mathrm{logs}^f_n(N_S)\right]\,,\\\nonumber
\Delta_n^{(2)}&=\frac{1}{N_T^2}\left\{v_n
-\frac{1}{N_S}\left[v_n+\delta v_n+\mathrm{logs}^v_n(N_S)\right]\right\}
+\frac{1}{N_S^2}\left\{w_n
-\frac{1}{N_S}\left[w_n+\delta w_n+\mathrm{logs}^w_n(N_S)\right]\right\}
\,.
\end{align}
For sufficiently large $N_T$, the correction $\Delta_n^{(1)}$ dominates over $\Delta_n^{(2)}$, which comprises the leading
$\mathcal{O}(1/N_T^2,1/N_S^2)$
lattice artefacts. In contrast to the $1/N_T^2$ correction
terms, the $1/N_S^2$ corrections do not have a significant
effect on our fits. Therefore, we neglect the terms containing $w_n$
and $\delta w_n$.

The $\Delta_n^{(1)}$ contribution arises from interactions 
with mirror charges on lattice replica~\cite{Trottier:2001vj}, 
which result in an effective static potential between
charges that are separated
by distances $aN_S$, but without self-energies.
Consequently, the high order behaviour of the
coefficients $c_n$ and $f_n$ depends on the very same renormalon. 
Precisely for this reason one needs to disentangle $\delta m$
from the $1/(aN_S)$ correction carefully,
in order to isolate the $d=1$ renormalon.

\begin{figure}[t!]
\centerline{\includegraphics[clip,width=.98\textwidth]{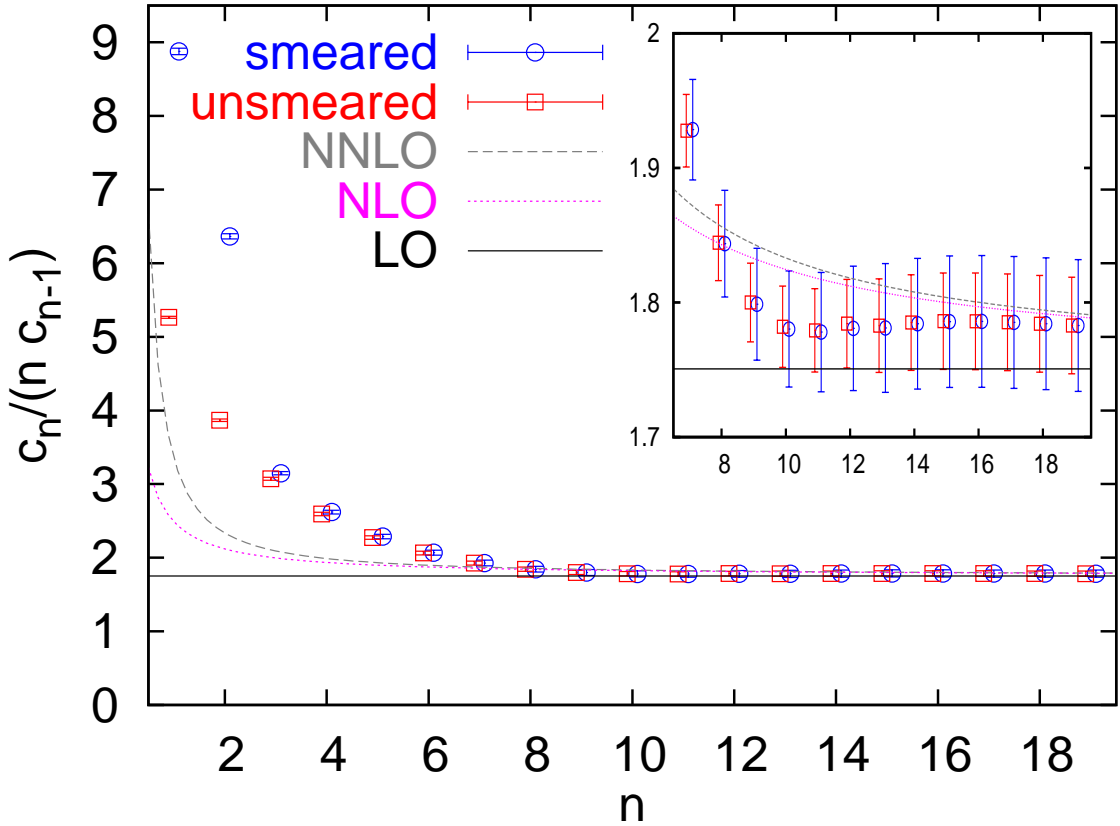}}
\caption{The ratios
$c_n/(nc_{n-1})$ for the smeared and unsmeared fundamental static
self-energies, compared to the prediction
eq.~(\protect\ref{cnratio}) at different orders of the
$1/n$ expansion.
\label{n20}}\end{figure}

The log-terms in $\Delta_n^{(1,2)}$ are due to the running of $\alpha$ according to the $\beta$-function. 
In the lattice scheme the first three $\beta$-function
coefficients $\beta_0$, $\beta_1$ and $\beta_2$ have been
calculated~\cite{Bode:2001uz}. The lack of knowledge of higher $\beta_j$ is leviated by the fact 
that these do not multiply the
leading $f_n, f_{n-1}, f_{n-2}, f_{n-3}$ coefficients: at sufficiently high $n$, 
the $f_n$ grow factorially and therefore subleading terms become
strongly suppressed. 

Eq. (\ref{eq:dlo}) is used for a global fit to all orders and all data from all our lattice volumes (see Table~I of \cite{Bauer:2011ws}). 
To illustrate the adequacy of the ansatz, in fig.~\ref{fig:fit}
we compare the fit to
the unsmeared triplet data,
for the case of $n=9$.
Note that the ansatz requires only four parameters per order, yet it results in reasonable values $\chi^2/N_{\mathrm{DF}}\approx 1.29$ and 1.46
 for smeared and unsmeared triplet data, respectively.

Our main finding is shown in fig.~\ref{n20}, where we confront the infinite volume extrapolated data for $c_n/(nc_{n-1})$ with
the theoretical prediction, eq. (\ref{cnratio}). Since smearing is a local operation that has little effect on long-range physics, 
the smeared and unsmeared data should exhibit the same
large $n$ behaviour, that is
dictated by the infrared renormalon. 
This universality is supported by our data, as~fig.~\ref{n20} illustrates. 
The octet representation data are also fully compatible with these findings, as we will detail in~\cite{BBP}.

\FIGURE{
\includegraphics[clip,angle=270,width=.88\textwidth]{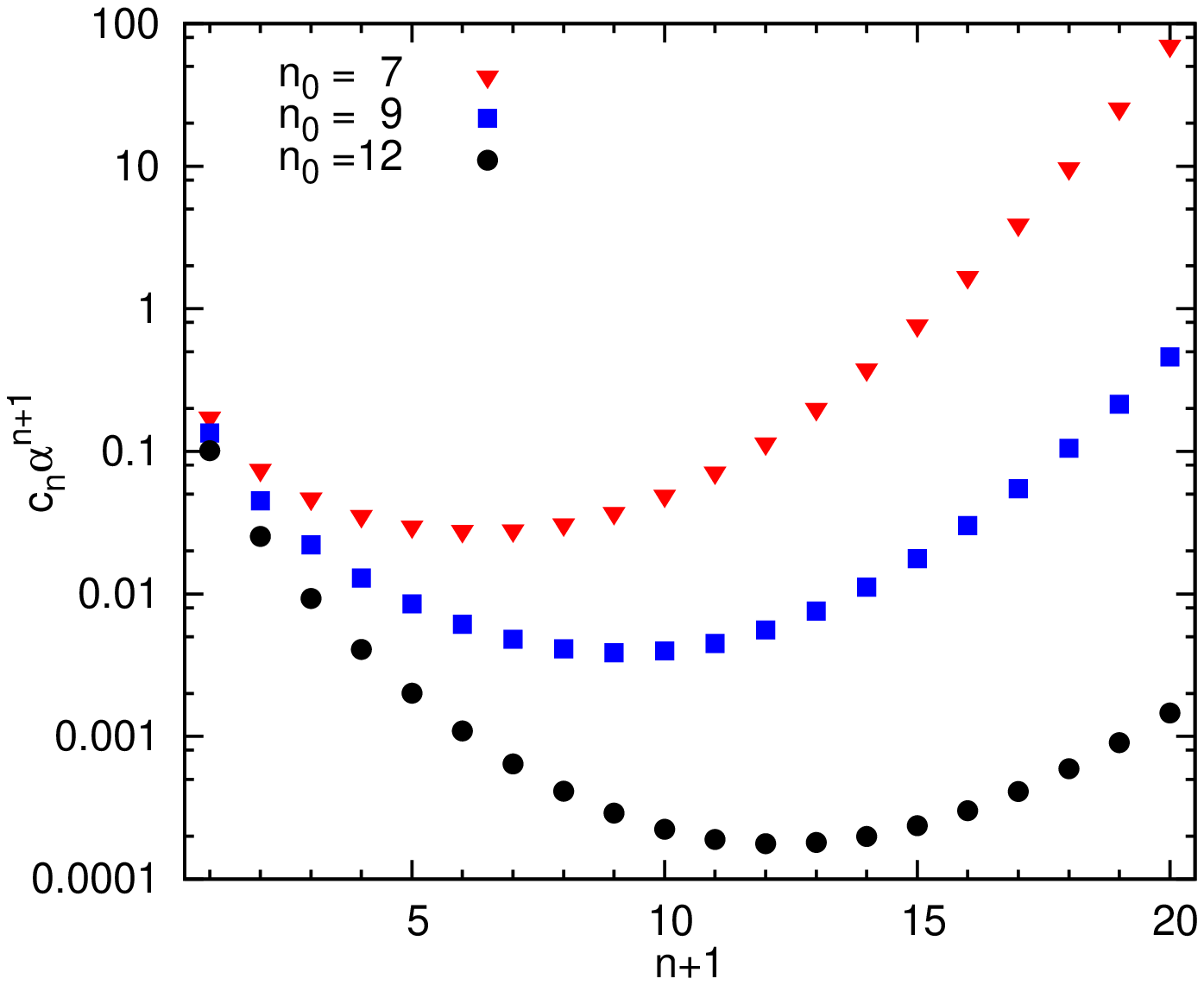}
\caption{Contributions to the sum eq. (\ref{deltam}) for three different values of $\alpha = 1/(|a_d|n_0)$, chosen such that $n_0 = 7,\,9$ and 12.
\label{diverge}}
}
In order to visualize the asymptotic behaviour of the perturbative expansion, 
in fig.~\ref{diverge} we display the order-by-order contributions to the sum eq. (\ref{deltam}). 
For $n_0= 7,\,9$ and 12, three different values of $\alpha=1/(|a_d|n_0)$
are chosen
(note that $n_0=7$, corresponding to $\beta=3/(2\pi\alpha)\approx 5.85$ or
$a^{-1}\approx 1.6$~GeV, is within the range covered by
non-perturbative lattice simulations). 
The contributions indeed become minimal at orders close to the
predicted values $n_0\approx n+1$. 

With the infinite volume coefficients~$c_n$ at hand, we can compute the normalization of the pole mass renormalon,
see eq.~(\ref{generalm}). For smeared triplet data we get 
$N^{\mathrm{lat}}_m=18.6(4)$, compared to $N^{\mathrm{lat}}_m=19.0(3)$ for the unsmeared case. 
As $\Lambda_{\MS}\,N^{\MS}_m=\Lambda_{\mathrm{lat}}\,N^{\mathrm{lat}}_m$, this translates to $0.65(2)$ in the $\MS$ scheme. 
This agrees remarkably well with the estimate, $N^{\MS}_m\approx 0.62$~\cite{Pineda:2002se,Lee}, 
from an expansion in the $\MS$ scheme up to $\mathcal{O}(\alpha^3)$.
Hence, the $d=1$ renormalon manifests itself in a lower bound of $\sim 0.65\,\Lambda_{\MS}$ for the precision with which one can determine the heavy quark pole mass.

\section{Summary}
Within the framework of NSPT, we have computed the static self-energy
of $\mathrm{SU(3)}$ gauge theory in four spacetime dimensions
to $\mathcal{O}(\alpha^{20})$ in the lattice scheme. 
Simulations on various lattice volumes and a careful treatment of finite size effects are vital for our main finding, the onset of a factorial growth of the
self-energy coefficients at an order $n\approx 9$, in agreement with the
theoretical renormalon prediction.

\begin{acknowledgments}
We benefited from discussions with 
V.\ Braun, F.\ Di Renzo, M.\ Garc\'{\i}a P\'erez,
H.\ Perlt, A.\ Schiller and C.\ Torrero.
The simulations were carried out on Regensburg's Athene HPC cluster
and at the Leibniz Supercomputing Centre in Munich.
C.B.\ acknowledges support by the Studienstiftung des deutschen
Volkes and by the Daimler und Benz Stiftung.
This work was supported by DAAD (Acciones Integradas
Hispano-Alemanas D/07/13355),
DFG SFB/TR 55, the
EU ITN STRONGnet grant 238353, the Spanish 
grant FPA2010-16963, and the Catalan grant SGR2009-00894.
\end{acknowledgments}

\end{document}